\documentclass[fleqn,twoside]{article}
\usepackage{espcrc2}
\usepackage{graphicx}
\usepackage[figuresright]{rotating}
\usepackage{amsmath,amsthm,amsfonts,amscd} 
\usepackage{eucal} 	 	
\usepackage{verbatim}      	
\usepackage{makeidx}       	
\usepackage{citesort}         	%
\usepackage{url}		
\usepackage{subfigure}
\title{The EHE Neutrino Search Capability of the IceCube Observatory}
\author{Aya Ishihara 
  \address[WISC]{Physics Department, University of Wisconsin, Madison WI, 53706, USA}%
  \thanks{Present address: Dept. of Physics, Chiba University, Chiba 263-8522 Japan}
  for the IceCube Collaboration}
\theoremstyle{plain} 

\theoremstyle{definition}

\theoremstyle{remark}

\begin{document}
\begin{abstract}

 An initial study of the extremely high energy (EHE) physics capability
 of the IceCube neutrino observatory is demonstrated by considering a
 GZK mechanism neutrino production model, which is a guaranteed source
 for EHE neutrinos. We study EHE event properties in the energy range  
 $10^{5}$ $<$ E $<$ $10^{11}$ GeV observed by IceCube using detailed
 Monte Carlo simulation.
 Results of a simulation study show that about $0.7$ EHE neutrino 
 events per year is expected with the full 
 IceCube configuration over a $0.03$ atmospheric muon background 
 which passed an EHE event criteria.
 It is also shown that with the
 present partial IceCube detector with the same criteria is capable of
 studying EHE physics and the sensitivity improves with the number of
 deployed strings in the ice.
 
 \vspace{1pc}
\end{abstract}
\maketitle
\section{\label{sec:intro}Introduction}

It is well known that there exists extremely high energy (EHE)
particles in the universe with energies approaching up to 
$\sim 10^{20}$ eV~\cite{nagano01}.
These EHE cosmic rays (EHECRs) may produce
neutrinos by various mechanisms, namely 
when they interact with surrounding matter or photon fields. 
The neutrinos are generated by the decays
of $\pi$ mesons via 
$\pi^{\pm}\to \mu^{\pm}\nu_\mu\to e^{\pm}\nu_e\nu_\mu$ process.
In the EHE region, collisions of EHECRs and the cosmic microwave
background photons produce cosmogenic neutrinos~\cite{berezinsky69}, a
consequence of a process known as the Greisen-Zatsepin-Kuzmin (GZK)
mechanism~\cite{GZK}. 

In general neutrinos are unique probes for exploring the high energy universe
mainly because of two distinct features. 
The first feature is that the existence of cosmic neutrinos implies that
there are hadron beams because only high energy {\it hadronic} processes
can produce neutrinos. 
The second feature is that neutrinos can penetrate dense matter and
radiation fields because of their small cross-sections, and therefore
can propagate over cosmological distances.
This implies that by observing neutrinos, one is able to learn about
stars and galaxies which are surrounded by dense gases or energetic
objects at very large distances. Such information is obscured when
observing with photons (e.g., radio, visible light, X-ray and
$\gamma$-ray). 
Because of these features, furthermore,
detection of EHE neutrinos
may shed light onto one of the
most difficult questions in modern astrophysics, how and where are
EHECRs produced. 

It is generally very difficult to detect high energy neutrinos
because one needs a large target volume for neutrinos to interact and
produce a detectable signal in a reasonable time scale.  
The IceCube neutrino observatory, however, 
provides a rare opportunity to overcome this difficulty.
Its km$^3$ fiducial volume which uses clean glacier ice deep below the
surface at the South Pole is a powerful tool to search for EHE
neutrinos. 
In this paper we discuss the capability of the IceCube experiment to
detect EHE neutrinos. 

The paper is outlined as follows:
In the next section, the overview of
the IceCube observatory and its present status are briefly mentioned.
The section~\ref{sec:ehe_icecube} describes the 
characteristics of EHE neutrino events in the IceCube detector.
We explain signals and backgrounds in our EHE neutrino search
and a possible strategy to discriminate neutrino events by 
considering their event topologies. In the section~\ref{sec:mc_results},
the main results from the detailed Monte Carlo simulation
study are shown, referring to the preliminary numbers concerning
effective area and event rate for cosmogenic EHE neutrinos.
We summarize our conclusion and future prospects
at the end.

\section{\label{sec:iceCube}The IceCube Neutrino Observatory}

IceCube is a next-generation cubic-kilometer scale
high energy cosmic neutrino telescope currently
under construction and in operation
at the geographic South Pole.
It uses 3 km thick glacial ice as 
a Cherenkov medium. Cherenkov photons emitted
from relativistic charged particles such as muons
are received by an array of Digital Optical Modules
(DOMs) which amplify and digitally sample the high-speed
photomultiplier tube (PMT) pulses {\it in situ}.
Each DOM encloses 10'' R7081-02 PMT made by Hamamatsu
Photonics in a transparent pressure sphere along
with the high voltage system, a LED flasher board for
optical calibration in ice, and a digital
readout board. The deep-ice DOMs are deployed
along electrical cable bundles which carry power and
information between the DOMs and surface electronics.
The cable assemblies, often called strings, are dropped
into holes drilled to a depth of 2450 meters. 
The DOMs occupy the last 1000 meters at intervals of 
17 meters where the glacial ice 
is transparent. DOMs are also frozen into
tanks located at the surface near the top of each hole
which constitutes an air shower array called IceTop.
IceTop provides us with the capability
to study the atmospheric muon background reliably. This feature plays
a key role in the IceCube EHE neutrino search as we will mention later.
The strings and tanks are arranged in a hexagonal lattice pattern
with a spacing of approximately 125 meters. At completion (planned
to be 2011), the array will comprise 4200 in-ice DOMs on 70 strings
and 320 modules in the surface array. Currently, IceCube 
includes 9 in-ice strings (540 DOMs) and 32 IceTop tanks (68 DOMs).

Events recorded by IceCube are generally categorized 
by geometrical patterns of photon distributions from the minimum
ionizing in-ice charged particles.
The ``track'' events are initiated either from 
penetrating neutrino-induced muons or taus, or from muons coming from
extensive air shower cascades above the surface {\it i.e.,}
the atmospheric muons.
Directional reconstruction of these tracks suppresses the atmospheric
muon background. By selecting neutrino events that are up going, this
effectively eliminates the atmospheric muon background since they are
down going. 
The other event pattern known as a ``cascade'' is induced by an
electromagnetic (EM) cascade via the $\nu_e$ charged-current interaction
or hadronic cascades via the neutral-current interaction of all neutrino
flavors. 
Those cascades generated inside IceCube detection volume emit spherical
shower-like patterns of Cherenkov light. 
Because muon and tau tracks have a longer propagation length in matter/ice 
than that of the electron and hadron cascades, tracks generally have a
larger effective area. 

EHE events that are expected to be seen by the IceCube observatory
can be considered beloinging to neither class, however, because of
difference in their major energy-loss processes. 
At the highest energies, photon emission patterns differ and their 
remarkably sizable energy deposit in the IceCube array 
may create a different characteristic signature. We describe this EHE
event topology in the next section. 

\section{\label{sec:ehe_icecube}EHE neutrino events in IceCube}
In the EHE region, 
because of the increase of the neutrino cross-section with energy,
neutrinos are more likely to be involved in interactions with matter 
during their propagation than to penetrate through the Earth.
Charged leptons and hadrons are generated in these interactions and 
the secondary produced
$\mu$'s and $\tau$'s travel the Earth 
losing their energies by 
undergoing many radiative reactions, {\it i.e.} 
EM cascades generated by $e^{\pm}$ pair creation, 
Bremsstrahlung, 
and hadronic cascades generated by the photonuclear interactions.
The IceCube detector is to observe these secondary $\mu$'s
and $\tau$'s as a main detectable channel 
of EHE signals~\cite{yoshida04}. 
There are two prominent
characteristics for these EHE neutrino events. One is that
a major fraction of them arrive at the detector
with a down-going geometry because the mean free path of EHE neutrinos
is much shorter than the typical path length in the Earth. 
Another is that the muon/tau
``track'' is accompanied with many ``cascades'' originating
from the various radiative energy loss processes.
The first characteristic makes it difficult to discriminate
between the atmospheric muon background events and neutrino induced events 
by using the Earth as a filter.
%
The second point, however,
provides another way to distinguish the EHE signals from background because 
the energy loss due to stochastic radiative processes is proportional to
the energy of muon(tau).  

\begin{figure}[tb]
\begin{center}
\includegraphics[width=8cm]{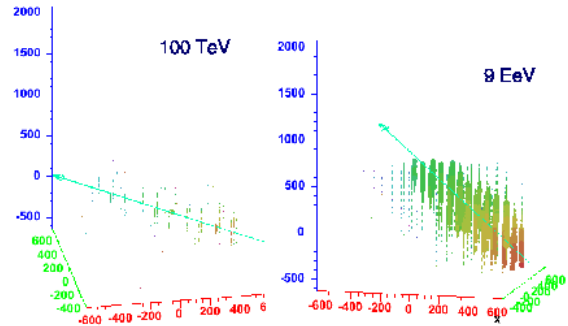}
\end{center}
\vspace{-1.2cm}
\caption{Examples of simulated IceCube events. Left panel shows
a 100 TeV muon track representing the conventional IceCube event
while the right panel indicates a 9 EeV muon EHE event.
Colored circles denote DOMs with more than one photon,
the size of the circles represents the number of photons and
the color indicate the first photon arrival time at each DOMs.
x, y and z axes are distances from the central position of the 
IceCube detector.
}
\label{fig:iceCube_event}
\end{figure}

A measurement of the energy deposit inside
the IceCube detection volume, therefore, leads to an estimation
of the track energy. 
The estimated track energy tells if it is of cosmic origin. This is
because the  expected spectra of secondary $\mu$'s and $\tau$'s
generated from the GZK neutrinos is much harder than that of atmospheric
muons~\cite{yoshida04}, and the measured energy (or its indicator)
should be able to exclude the atmospheric muon events in a relatively
straightforward manner.  
Figure~\ref{fig:iceCube_event}
shows examples of simulated IceCube events at different energies. 
One can see
that the EHE muon radiates 
a large number of Cherenkov photons.
The amount of Cherenkov photons in an event,
the ``brightness'' of the event in other words, appears
to indicate the track energy. 
Detailed Monte Carlo study has shown that
the number of detected photons is a robust energy indicator
to search for EHE signals, which is described in the following section.

\section{\label{sec:mc_results} Results}
\subsection{Simulation Setup and Signal Domain Criteria}

Transportation of the EHE neutrinos and their secondary particles
from the Earth's surface to the detector depth is calculated by the 
JULIeT package \cite{juliet} as described in Ref.~\cite{yoshida04}.
The IceCube Monte Carlo simulation package then generates events with
energies and intensities following the obtained fluxes.
The primary cosmic neutrino flux in this paper
is assumed to be GZK cosmogenic neutrinos as 
calculated in Ref.~\cite{yoshida93}.
For the atmospheric muon flux, which is considered our main background,
we take the analytically fitted result using a Corsika
simulation~\cite{Corsika} assuming an initial cosmic-ray proton flux of 
the form of $E^{-3}$.
We should note that the atmospheric muon background in the EHE regime
is highly uncertain due to our poor knowledge of
the EHECR mass composition and the muon bundle intensity.
The background intensities will be deduced from the IceCube data in the future.
The present simulation
chain considers events induced by the charged leptons that interact within 860m from the center of detector because the secondary produced muons and taus are our main EHE events in IceCube~\cite{yoshida04}. 
The energy range under consideration is
between $10^5$ and $10^{11}$ GeV near the detector. 

\begin{figure}[hbt]
\begin{center}
\includegraphics[height=40mm]
{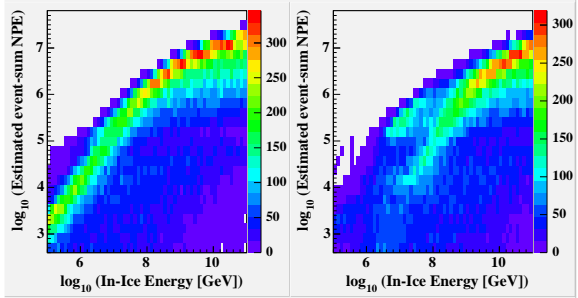}
\end{center}
\vspace{-1.2cm}
\caption{Simulated event distributions on a plane of
NPE and the charged lepton energy. 
Energy plotted here is the value
when the incoming particle is at the surface of IceCube
detection volume which is defined as a sphere of 860 meters in radius 
in the present study. 
The left plot shows muon events and the right plot shows tau events.
A suppression of energy loss for taus compared to that of muons 
and the contributions from tau-decays are visible in the right plot. 
}

\label{fig:EnergyNPE}
\end{figure}
As described in Section~\ref{sec:ehe_icecube},
the total amount of Cherenkov light detected for an event
is expected to be sensitive to the energy of incoming EHE muon or tau.
The IceCube detector can measure the Cherenkov luminosity per event as
the event-total number of photo-electrons (NPE) 
detected by each DOM. 
Figure~\ref{fig:EnergyNPE} shows
distributions of NPE as a function of the muon and tau energy.
It shows a correlation between NPE and the muon/tau energy going through
the IceCube volume. 
Notice that the track geometry also affects the observed NPEs since 
a track passing further away from the DOMs will result in less Cherenkov light within the IceCube detector.
A distinct feature in the plots is that the primary energy of the incoming leptons is still a deciding factor in the observed NPEs, 
which implies that NPE
is a robust indicator of the lepton energy without having to rely on event geometry
reconstruction. 

\begin{figure}[t]
\begin{center}
\includegraphics[height=40mm]{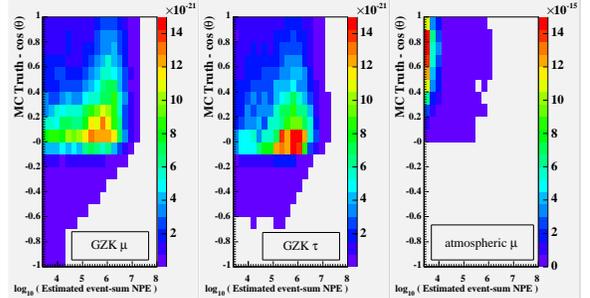}
\end{center}
\vspace{-1.2cm}
\caption{
Event distribution on the plane of NPE and cosine of zenith angle
obtained by Monte Carlo simulation.
Plotted in the left and middle are that of GZK neutrino-induced 
muon and tau signals respectively and the background atmospheric muons on the right.
}
\label{fig:NPECosTheta}
\end{figure}

The zenith angle distributions are quite different for 
muon and taus originating from the propagation of cosmogenic GZK neutrinos and 
atmospheric muons, however. 
As down-going GZK muons and taus have a higher tendency to arrive from near the horizontal direction than
the atmospheric background~\cite{yoshida04},
one can expect a difference in the distribution of 
zenith angles and NPE for EHE signals and background.
Plotted in Figure~\ref{fig:NPECosTheta} are
the simulated event distributions on the plane of NPE and zenith
angle of muons and taus expected from the GZK model and the atmospheric muon backgrounds,
which behave consistently with our expectation. 
Taking advantage of such differences, we introduce 
the signal domain on the NPE-Zenith angle plane where the GZK events
dominates over the background defined as 
\begin{equation}
{\rm NPE}\geq \Bigl\{
\begin{tabular}{cc}
$5\times 10^5$ & if $\cos\theta\geq$ 0.1, \\
$1\times 10^5$ & otherwise.
\end{tabular}\label{eq:eheselection}
\end{equation}
The expected event rate and effective area with this criteria
are discussed next.

\subsection{Event Rate}

\begin{figure}[htb]
  \begin{center}
    \begin{tabular}{c}
      \subfigure{
	\includegraphics[width=70mm]{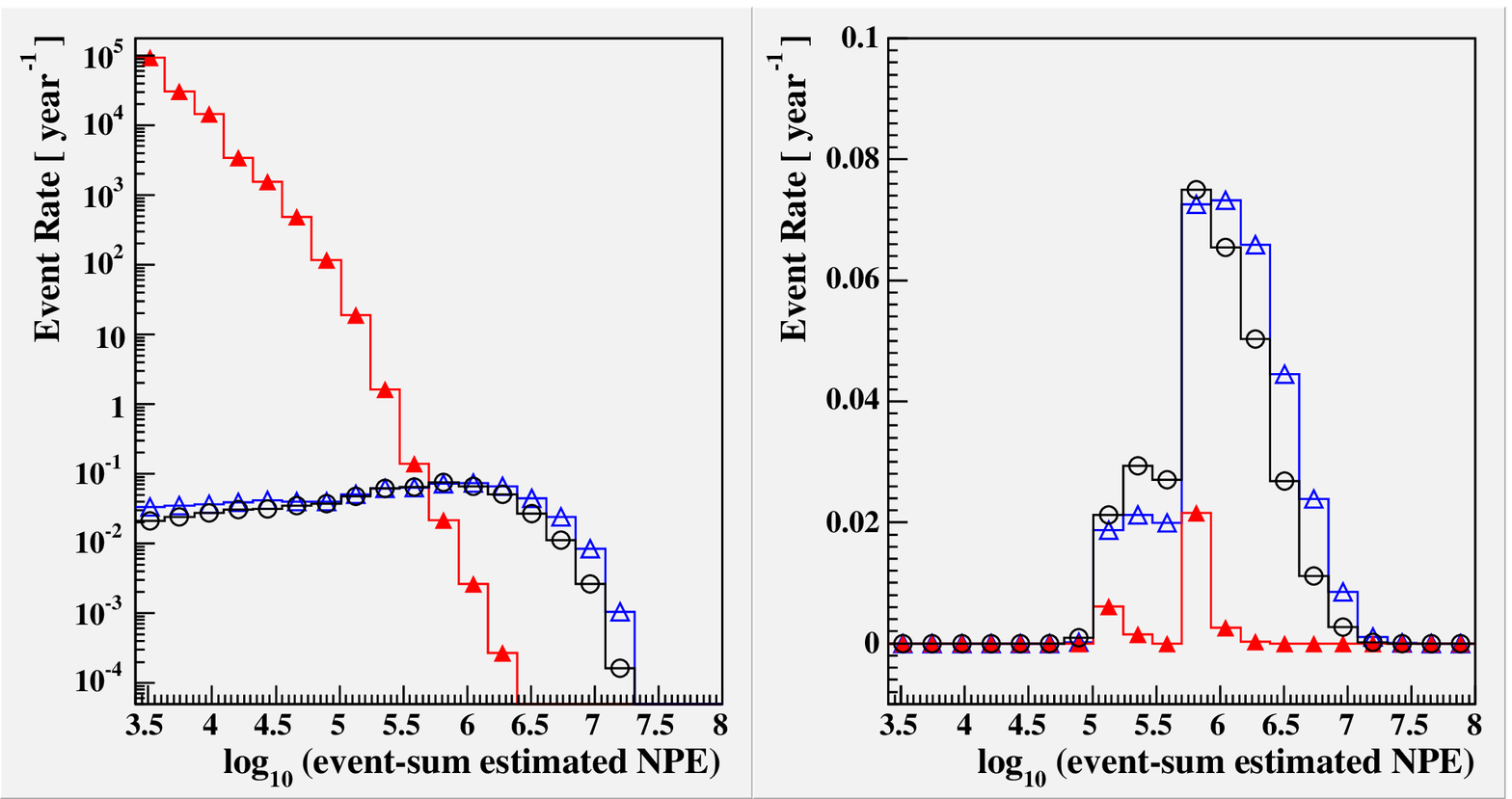}
      }\\
      \subfigure{
	\includegraphics[width=70mm]{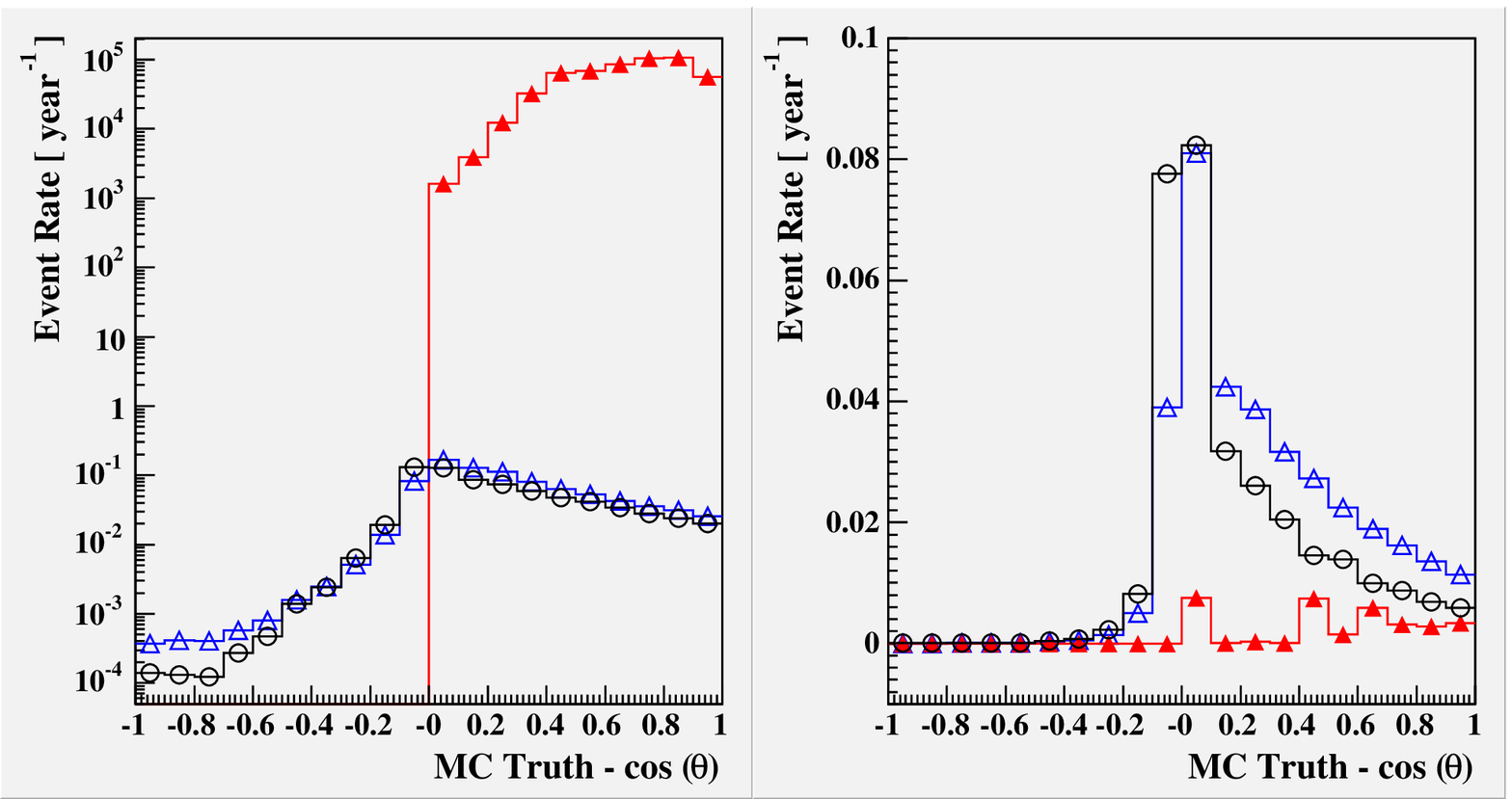}
      }
    \end{tabular}
    \vspace{-1.2cm}
    \caption[]{
      $\mu/\tau$ GZK signal and atmospheric $\mu$ event rate as 
      function of NPE (upper) and $\cos (\theta)$ (lower).
      Blue and black markers indicate GZK $\mu$, 
      $\tau$ respectively and red markers indicate atmospheric $\mu$. 
      Left plots display event rates without 
      background cut and right plots show after the 
      signal domain selection. Multiple muon events are not included in
   the atmospheric muon flux.}
    \label{fig:eventrate}
  \end{center}
\end{figure}

The expected signal and background event rate are shown 
in Figure \ref{fig:eventrate} as functions of NPE and zenith angle.

Major parts of the dominant atmospheric background in the left plots are
rejected by the selection criteria defined as Eq. (\ref{eq:eheselection})
in the right plots.  With these selection criterion, the GZK muon and tau signal events dominate.
It is also found that the number of tau events are comparable to or even
greater than that of muon events. In the EHE regime, the tau does not decay
but is subject to radiative energy loss processes.
Its heavier mass suppresses the energy loss compared to
muons with the same energy, which gives a higher
survival probability from near the horizontal direction. 

Shown in Table~\ref{table:eventrate} are the event rates from 
different string numbers deployed in ice.
The same selection criteria are applied for all the 
string configurations, including the present 9 string setup.
Event rate increases almost constantly as more 
strings are installed.

\begin{table}[htb]
\small
\caption{The preliminary IceCube EHE event rate for different 
deployed in-ice string configurations. Muons and taus produced
from the GZK model and the muon atmospheric background are shown.}
\label{table:eventrate}
\vspace{5mm}
\newcommand{\m}{\hphantom{$-$}}
\newcommand{\cc}[1]{\multicolumn{1}{c}{#1}}
\begin{tabular}{@{}llllll}
\hline
{string number}       & \cc{$9$} & \cc{$20$} & \cc{$40$} & \cc{$60$} & \cc{$80$} \\
\hline
GZK $\mu$                & $0.067$ & $0.12$  & $0.21$  & $0.28$  & $0.35$  \\
GZK $\tau$               & $0.055$ & $0.11$  & $0.19$  & $0.25$  & $0.31$  \\
atmospheric $\mu$        & $0.009$ & $0.010$ & $0.012$ & $0.025$ & $0.033$ \\

\hline
\end{tabular}
\end{table}

\subsection{Effective Area}

\begin{figure}[htb]
\begin{center}
\includegraphics[height=40mm]{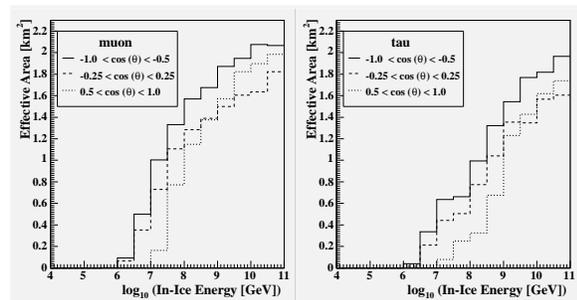}
\end{center}
\vspace{-1.2cm}
\caption{Muon (left) and tau (right) effective area as a function of incoming track energy 
  with the 80 string configuration based on the EHE signal criteria. 
 Dashed line denotes horizontal events and dotted line represents down-going events.
 Solid line indicates up-going events, although the probability of having EHE
 neutrinos with up-going geometry is very low, because of the Earth's
 sheltering effect.  
}
\label{fig:EffectiveArea}
\end{figure}

The effective area resulted from the EHE signal domain cut
is plotted in Figure~\ref{fig:EffectiveArea} as a function
of incoming charged lepton energy for different zenith angles. 
%
%
The NPE-based EHE event selection criteria decreases
effective area in lower energy region while it enhances
the detection efficiency for events with energies
above $10^8$ GeV. These events are so luminous 
that the DOMs receive many photons
even from the tracks propagating far outside the instrumented detector
volume. 
The area reaches $\sim$ 2 km$^2$,
a factor of two larger than the IceCube physical cross section
in energy region above $10^{10}$ GeV.
The area for up-going geometry is slightly larger
as the NPE cut is loose for those geometries.

\section{Summary and Future Prospective}

We have considered the capability of the IceCube detector in the 
search for EHE neutrinos without extensive energy reconstruction.
The first level background rejection and signal selection is performed 
in terms of NPE and MC truth incoming direction. With this selection criteria, it is shown by a Monte Carlo simulation 
that about $0.7$ EHE GZK events per year is expected with the full 
IceCube configuration over a background of $0.03$ atmospheric muons.
 
The expected GZK event rate with the present 9 string configuration is about 
$0.12$ events per year with the same event selection.
The number increases with the number of strings deployed from 9 strings 
to the full number of strings.


The study presented in this paper has not involved energy and 
geometry reconstruction of muon and tau tracks.
We expect that good geometry reconstructions will support our results
based on the Monte Carlo truth angle information.
In addition, utilizing energy reconstruction should lead to a
significant improvement of detector performance, 
especially for uncontained 
events that traverse outside the IceCube instrumentation volume.
Ongoing development of EHE reconstruction methods 
will improve our sensitivity to EHE cosmic neutrinos in near future.

The results we have shown here heavily rely on NPE-Energy relations.
These contain sizable uncertainties due to our incomplete understanding
of the ice properties and detector responses to high NPE signals.
The ice properties include the optical properties of the glacier ice and the behavior of the Cherenkov photon
propagation in the hole ice which was melted and refrozen 
during the detector deployment.
In order to reduce the systematic errors in the present study,
the absolutely calibrated light source called the standard candle (SC)
has been deployed in neighborhood of the DOMs in ice.
The SC consists of a nitrogen laser, the calibrated PMT, and the related optics
system and is able to emit pulses which are approximately equivalent to a
PeV EM cascade. 
Analyzing the data generated by the SC should reduce current
uncertainties and improve the reliability of the EHE neutrino search.  

The intensity of the atmospheric muon background in the EHE energy region 
is not well understood as we mentioned in Section~\ref{sec:mc_results},
and we have assumed the Corsika-based analytical model 
in the present simulation study.
In the future, however, it is expected that reasonable statistics of the data with NPE below 
$\sim 10^5$, allows us to build a reasonable empirical model to predict
number of the background events in the EHE signal domain.
We also would like to point out that in any case our sensitivity would not be changed significantly
by varying the definition of energy based background rejection factor
because the GZK neutrino induced muons and taus exhibit
an extremely hard energy distribution in contrast to the very soft ($\sim {\rm E}^{-4}$)
spectrum of the atmospheric muons.
Besides, the atmospheric muon events can be {\it experimentally}
identified and excluded by looking for coincidence events between
the deep-ice IceCube strings and the IceTop surface array.
With this power of the IceTop air shower array, the hybrid analysis
using the in-ice DOMs and the surface DOMs, even with a limited data
statistics, provides us further information about the background,
including events consist of multiple muons, in the very high energy
region. 

\section{Acknowledgments}

We acknowledge the support from the following agencies:
National Science Foundation-Office of Polar Program,
National Science Foundation-Physics Division,
University of Wisconsin Alumni Research Foundation,
Department of Energy, and National Energy Research Scientific Computing
Center
(supported by the Office of Energy Research of the Department of Energy),
the NSF-supported TeraGrid system at the San Diego Supercomputer Center
(SDSC),
and the National Center for Supercomputing Applications (NCSA);
MEXT (Ministry of Education, 
Culture, Sports, Science, and Technology) in Japan;
Swedish Research Council,
Swedish Polar Research Secretariat,
and Knut and Alice Wallenberg Foundation, Sweden;
German Ministry for Education and Research,
Deutsche Forschungsgemeinschaft (DFG), Germany;
Fund for Scientific Research (FNRS-FWO),
Flanders Institute to encourage scientific and technological research in
industry (IWT),
Belgian Federal Office for Scientific, Technical and Cultural affairs
(OSTC);
the Netherlands Organisation for Scientific Research (NWO);
M.~Ribordy acknowledges the support of the SNF (Switzerland);
J.~D.~Zornoza acknowledges the Marie Curie OIF Program (contract 007921).


\end{document}